# Impact of the Design Parameters on the Microwave Displacement Sensor Performance

Premsai Regalla and A. V. Praveen Kumar
Department of Electrical and Electronics Engineering, Birla Institute of Technology and Science (BITS)
Pilani campus, Rajasthan-333031, India.
Premsairegalla999@gmail.com, Praveen.kumar@pilani.bits-pilani.ac.in

*Abstract*— The investigations have been conducted on the involved design parameters to analyze the behavior of the microwave displacement sensor output characteristics. To implement this, a dielectric resonator loaded to a reflection mode operated microstrip line circuit is proposed. For the proposed reflection mode sensor, the sensor features like resonant frequency, impedance matching position, sensitivity, and dynamic range are sensitive to the displacement of DR to microstrip line. The impact of the substrate shape and size, and resonators geometrical properties are numerically analyzed and experimentally validated by using VNA. The sensor analysis shows good matching between both HFSS simulations and VNA measurements.

## I. INTRODUCTION

Microwave sensors are low-cost, highly sensitive, and more accurate compared to the other frequency regions like low frequency sensors and/or high frequency sensors [1]. Consequently, an increased awareness towards microwave displacement sensors has been generated recently due to their small size, operating flexibility, and measuring techniques [2]. Most of these sensors use microwave resonators that are sensitive to displacement leading to change in the resonant characteristics of the circuit in which they are used. The most utilized microwave elements are either metallic or dielectric families [2]-[4]. All of these sensors follow one of the principles like shift-in frequency, split-in frequency, variable phase, or single frequency. Among various sensing principles used, fixed frequency-variable magnitude sensors are highly robust [5]-[6]. Additionally, they can be validated through multiple realization techniques [5]-[6]. The displacement sensor features such as sensitivity, linearity, and sensing range can be dependent on many parameters like resonating element, its feeding technique, mode of operation, and the selected output variable. Apparently, the involved design parameters like shape and size of ground and/or substrate may alter the sensor performance. So, it is essential to investigate on design parameters of displacement sensors which are essential in various industrial applications.

In this work, to study the effect of the design parameters on the sensor response, the authors proposed a reflection mode operated DR based displacement sensor. The involved design parameters are substrate and resonator, and the considered output sensor features are frequency, sensitivity, and range. This sensor is a single frequency operating system. The sensor design and its various aspects are investigated in Section. II. Sensor realization is presented in Section. III, followed by conclusion in Section. IV.

## II. SENSOR DESIGN AND ANALYSIS

Proposed reflection mode operated linear displacement sensor is graphically represented in Fig.1. The involved design parameters are substrate, and cylindrical dielectric resonator. The positional change in DR is represented as *dx*. The change in *dx* creates impedance mismatch, which causes variation in $|S_{11}|$. The change in $|S_{11}|$ for each *dx* is helpful in determining various sensor output characteristics like resonant frequency, optimum impedance matching position, sensitivity, and dynamic range. The above mentioned sensor features are highly dependent on the various factors like substrate relative permittivity, its shape and size, and resonators geometrical properties like diameter and aspect ratio. So, a detailed numerical investigation on these aspects is presented as follows.

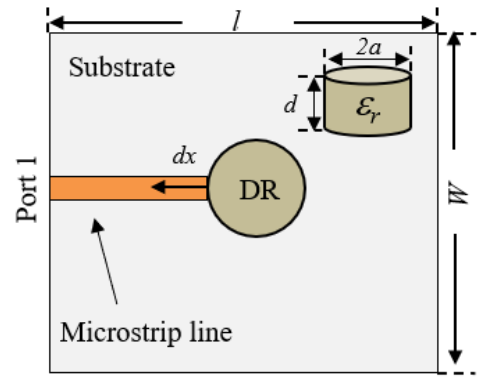

Fig.1. Graphical representation of the proposed sensor

### A) Impact of the substrate relative permittivity

To understand the impact of the substrate relative permittivity on the sensor performance, FR4 and RT/Duroid substrates are considered for the analysis. The considered DR dimensions are *2a*=19.43mm, *a/h*=1.33, $\varepsilon_r$=24. The substrate specifications are mentioned in Table 1. Initially, DR position is fixed at the substrate center, where it touches the microstrip line. The further movement of DR helps to find the optimum impedance matching position with minimum $|S_{11}|$. The frequency response of both the substrates at optimum impedance position is shown in Fig.2. and the corresponding sensitivity response at each *dx* for both the cases are shown in Fig. 3. In sensitivity curve, we can see pre-matching and post-matching regions. The 1st region helps for high sensitivity, and 2nd for wide range. The sensor performance parameters like resonant frequency, sensitivity, dynamic ranges are clearly mentioned in Table. 1.

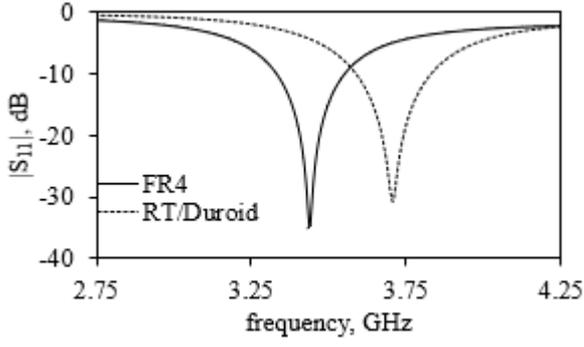

Fig.2. Simulated $S_{11}$ spectrum at optimum $dx$ position of the proposed sensor for both the substrates

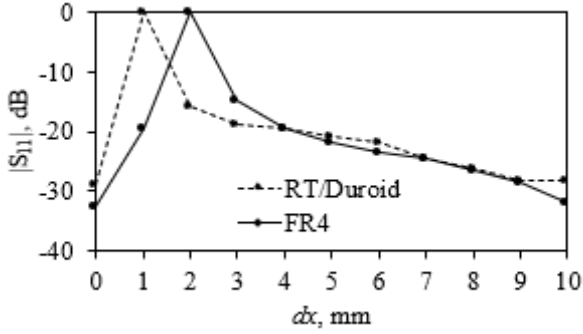

Fig.3. Sensitivity curve generated from Fig.2 at 3.34 GHz

Table 1: Impact of the substrate relative permittivity on the sensor performance

| Sensor characteristics | FR4 | RT duroid |
|---|---|---|
| Relative permittivity | 4 | 2.3 |
| Loss tangent | 0.02 | 0.001 |
| size | 115×115 mm² | 115×115 mm² |
| height | 1.6 mm | 1.57 mm |
| $f_r$ | 3.47 GHz | 3.74 GHz |
| min $|S_{11}|$, dB | 36 | 28 |
| Optimum $dx$ | 2 mm | 1 mm |
| sensitivity | 18 dB/mm | 28 dB/mm |
|  | 4.5 dB/mm | 3 dB/mm |
| Dynamic range | 0-2 mm | 0-1 mm |
|  | 2-10 mm | 1-10 mm |

*B) Impact of the substrate shape and size*

To analyse the impact of the substrate shape and size, we have considered circular and square shaped FR4 substrates with various sizes like 130×130 mm², 115×115 mm², 75×75 mm², and 50×50 mm². The frequency responses for each substrate size at optimum $dx$ position, where the min $|S_{11}|$ can be achieved as shown in Fig.4. (a) for square and (b) for circular shapes. Similarly, the corresponding sensitivity curves are shown in Fig.5. As per the HFSS analysis, the impact of the substrate shape and size on the various sensor output parameters are listed in Table. 2. As represented in the Table 2, the decrease in the substrate size results in decreasing the sensitivity with an increase in the optimum $dx$ position and dynamic range in highly sensitive region. Whereas the change in substrate shape is not showing much impact on the sensor performance. i.e., the sensor response for square and circular substrates are almost identical. For the practical convenience, focused only on the FR4 substrate.

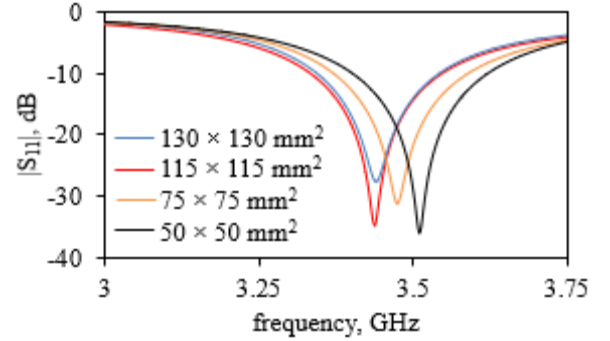

(a)

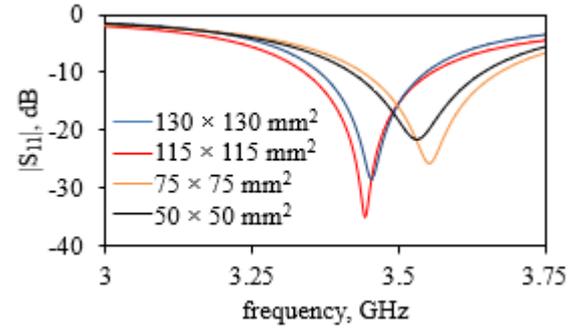

(b)

Fig.4. Simulated $|S_{11}|$ vs frequency curves for the proposed sensor at optimum $dx$. (a) square substrates, (b) circular substrates

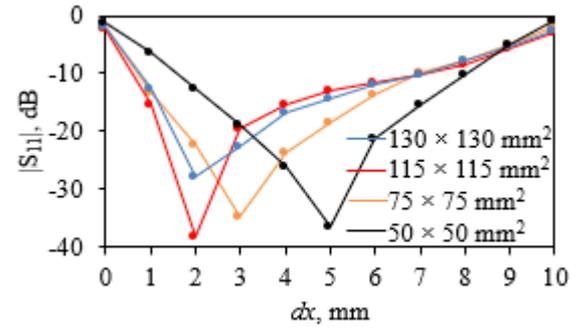

(a)

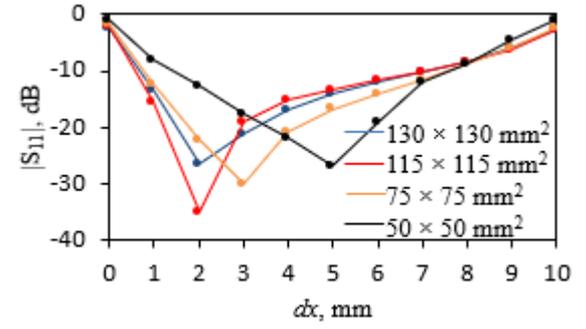

(b)

Fig.5. Simulated sensitivity curve generated from Fig.4 (a) square substrates, (b) circular substrates

Table 2: Impact of the substrate shape and size on the sensor performance

| Sensor characteristics | 130×130 mm² | 115×115 mm² | 75×75 mm² | 50×50 mm² |
|---|---|---|---|---|
| **Square substrate(s)** | | | | |
| $f_r$ (GHz) | 3.46 | 3.47 | 3.49 | 3.52 |
| min $|S_{11}|$, dB | 28 | 36 | 32 | 35 |
| Optimum $dx$ | 2 mm | 2 mm | 3 mm | 5 mm |
| Sensitivity (dB/mm) | 14 / 3.5 | 18 / 4.5 | 10 / 4.5 | 7 / 7 |
| Dynamic range (mm) | 0-2 / 2-10 | 0-2 / 2-10 | 0-3 / 3-10 | 0-5 / 5-10 |
| **Circular substrate(s)** | | | | |
| $f_r$ (GHz) | 3.47 | 3.48 | 3.52 | 3.55 |
| min $|S_{11}|$, dB | 28 | 36 | 32 | 28 |
| Optimum $dx$ | 2 mm | 2 mm | 3 mm | 5 mm |
| Sensitivity (dB/mm) | 14 / 3.5 | 18 / 4.5 | 10 / 4.5 | 5.6 / 5.6 |
| Dynamic range (mm) | 0-2 / 2-10 | 0-2 / 2-10 | 0-3 / 3-10 | 0-5 / 5-10 |

*C) Impact of the DR geometrical properties*

Here, the analysis has been done on the DR geometrical parameters to know its effect on the sensor performance. 115×115 mm² FR4 substrate size is considered, and the obtained outcomes are provided in Table 3. Initially, **Impact of the dielectric constant:** increase in dielectric constant ($\varepsilon_r$) leads to decrease in both the resonant frequency and range (only highly sensitive region), but sensitivity increases. Another, **Impact of the diameter:** increase in diameter ($2a$) leads to decrease in both the resonant frequency and sensitivity with an increase in range. Finally, **Impact of the aspect ratio:** increase in aspect ratio ($a/h$) leads to increase in both the resonant frequency and sensitivity with a decrease in range. Hence, the proper selection of DR is necessary to achieve the required performance.

Table 3: Impact of DR parameters on the sensor performance

| Sensor characteristics | $f_r$ (GHz) | min $|S_{11}|$ | Opt. $dx$ | Sensitivity (dB/mm) | Range (mm) |
|---|---|---|---|---|---|
| **Impact of the dielectric constant** | | | | | |
| 16 | 3.94 | 16.5 | 4 | 3.6 | 0-4 |
| 20 | 3.66 | 32 | 4 | 8 | 0-4 |
| *24* | 3.48 | 36 | 2 | 18 | 0-2 |
| **Impact of the diameter** | | | | | |
| 15.9 | 4.1 | 35 | 2 | 17.5 | 0-2 |
| 19.43 | 3.48 | 36 | 2 | 18 | 0-2 |
| *25* | 3 | 25 | 5 | 5 | 0-5 |
| **Impact of the aspect ratio** | | | | | |
| 0.67 | 2.8 | 13 | 4 | 2.8 | 0-4 |
| 1 | 3.14 | 48 | 4 | 11 | 0-4 |
| *1.33* | 3.48 | 36 | 2 | 18 | 0-2 |

*Only pre-matching region is mentioned for sensitivity and range

## III. MEASUREMENTS AND RESULTS

The fabricated sensor prototypes are shown in Fig.6. For the fair demonstration, by keeping the practical constraints in mind, two different sized circular and square shape FR4 structures are fabricated. Measurements are performed by using Vector Network Analyzer (VNA).

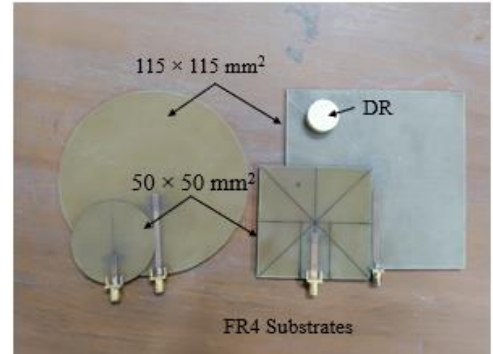

Fig.6 Fabricated prototype

The VNA measured frequency response of the fabricated prototypes for large (115×115 mm²) and small (50×50 mm²) substrates in both the shapes (circular and square) are shown in Fig.7. The measured resonant frequency for square shaped large substrate is 3.4 GHz and small one is 3.45 GHz. Similarly, 3.41 GHz and 3.46 GHz for circular substrates. For both the shape and sizes, the impedance matching position is identical to HFSS simulations. The sensitivity curves are shown in Fig.8 (a) for square and (b) for circular shape substrates. For large square substrate, the measured sensitivity curves exhibit 18.5 dB/mm in 0-2mm range and 4.7 dB/mm in 2-10 mm range. Similarly, for small square substrate, the measured sensitivity curves exhibit 6.7 dB/mm in 0-5mm range and 6.6 dB/mm in 5-10 mm range. In circular shape, for both the sizes, the measurement responses are identical to square substrate responses. Close approximation is observed between HFSS simulations and VNA measurements as shown in Fig.8.

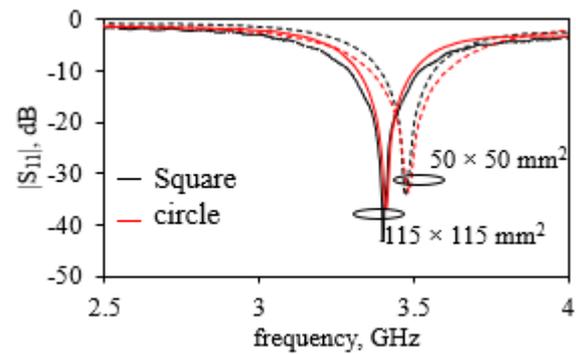

Fig.7. VNA measured (VNA) frequency response of the fabricated sensor for large and small substrates in both the shapes.

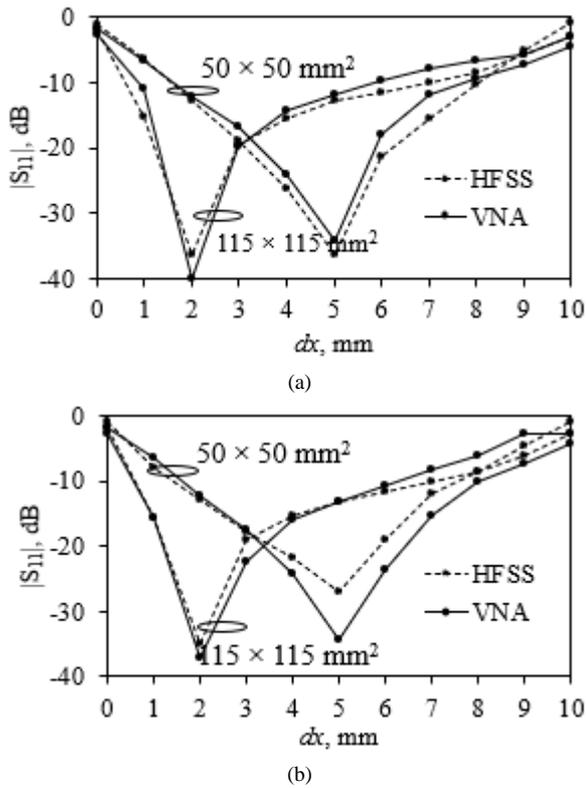

(a)

(b)

Fig.8. Comparison between the measured (VNA) and simulated (HFSS) sensitivity responses for (a) square shape (b) circular shape.

## IV. CONCLUSION

The investigations have been conducted on the microwave sensor design parameters to analyze the sensor performance. Different shape and sized substrates are realized for the proposed analysis feasibility. Successfully identified the role of each component in sensor design involvement for the desired sensor feature. This kind of single frequency operated sensors can be easily validated through low-cost reflectometer setups instead of VNA. The flexibility of multiple realization techniques is the merit for single frequency sensors.